\documentclass[conference]{IEEEtran}
\IEEEoverridecommandlockouts
\usepackage{cite}
\usepackage{amsmath,amssymb,amsfonts}
\usepackage{algorithmic}
\usepackage{graphicx}
\usepackage{comment}
\usepackage{booktabs}
\usepackage{hyperref}
\usepackage{textcomp}
\usepackage{xcolor}
\def\BibTeX{{\rm B\kern-.05em{\sc i\kern-.025em b}\kern-.08em
    T\kern-.1667em\lower.7ex\hbox{E}\kern-.125emX}}

\makeatletter
\newcommand{\linebreakand}{%
  \end{@IEEEauthorhalign}
  \hfill\mbox{}\par
  \mbox{}\hfill\begin{@IEEEauthorhalign}
}
\makeatother
    
\begin{document}

\title{Story2MIDI: Emotionally Aligned Music Generation from Text\\
}

\author{
\IEEEauthorblockN{Mohammad Shokri}
\IEEEauthorblockA{The Graduate Center, CUNY \\ New York, US \\ mshokri@gradcenter.cuny.edu}
\and
\IEEEauthorblockN{Alexandra C. Salem}
\IEEEauthorblockA{The Graduate Center, CUNY \\ New York, US \\ asalem1@gradcenter.cuny.edu}
\and
\IEEEauthorblockN{Gabriel Levine}
\IEEEauthorblockA{The Graduate Center, CUNY \\ New York, US \\ glevine@gradcenter.cuny.edu}
%
\linebreakand
\IEEEauthorblockN{Johanna Devaney}
\IEEEauthorblockA{Brooklyn College, CUNY \\ New York, US \\ johanna.devaney@brooklyn.cuny.edu}
\and
\IEEEauthorblockN{Sarah Ita Levitan}
\IEEEauthorblockA{Hunter College, CUNY \\ New York, US \\ slevitan@hunter.cuny.edu}
}

\maketitle

\begin{abstract}
In this paper, we introduce \textit{Story2MIDI}, a sequence-to-sequence Transformer-based model for generating emotion-aligned music from a given piece of text. To develop this model, we construct the Story2MIDI dataset by merging existing datasets for sentiment analysis from text and emotion classification in music. The resulting dataset contains pairs of text blurbs and music pieces that evoke the same emotions in the reader or listener. Despite the small scale of our dataset and limited computational resources, our results indicate that our model effectively learns emotion-relevant features in music and incorporates them into its generation process, producing samples with diverse emotional responses. We evaluate the generated outputs using objective musical metrics and a human listening study, confirming the model’s ability to capture intended emotional cues.
\end{abstract}

\section{Introduction}
We live in a world with an ever-growing demand for entertainment and multimedia content. The rise of social media and platforms for music, audio-books, and podcasts has gained tremendous momentum. At the heart of many of these forms of entertainment lies a narrative, a story that drives the experience, whether in a film, a game, a podcast, or a documentary. Narratives are powerful tools for evoking emotion. As audiences engage with a story, they often experience a dynamic emotional journey shaped by the characters \cite{vishnubhotla2024emotion,brahman2020modeling}, plot developments \cite{somasundaran2020emotion}, and underlying themes \cite{reagan2016emotional}. This emotional progression within a story, is integral to the impact that stories have on readers and viewers.
The emotional affect of a story is integral to the impact that it has on readers and viewers. Just like stories, music also has a remarkable impact on listeners’ emotional states \cite{juslin2011handbook} and is widely recognized as a means of expressing emotions \cite{forero2023words, davies1994musical}. Because of this emotional power, creative producers often accompany narratives with background music, which has been shown to significantly influence the audience’s emotional engagement and enhance the immersive quality of the content \cite{tan2007viewers, muller2024neural}. With the rapid progress of AI in various modalities, new models are introduced on a daily basis, capable of generating content in different modalities. 

The goal of our research is to develop a model capable of generating music that aligns with a given story, thereby enhancing its emotional impact. Narratives are complex, often guiding readers or viewers through evolving emotional arcs. Ideally, an effective model should be able to mirror this emotional trajectory and reinforce the story’s intended affective experience through music. However, this problem remains underexplored, in part due to the lack of large-scale datasets that pair narrative text with emotionally aligned music. 
As a first step towards our goal, this study focuses on generating music that captures the holistic emotional tone of a piece of text. In this study, we collect text blurbs from an existing sentiment analysis dataset annotated by humans. Using these sentences, we build a dataset of emotionally aligned story–music pairs. We then propose a Transformer-based encoder–decoder model \cite{vaswani2017attention} that generates music intended to evoke the same emotion as the input text. To ensure the model learns the structure of symbolic music, we pre-train the decoder on a large-scale symbolic music dataset before fine-tuning it on our emotion-aligned data.

\section{Related Work}

Affective Music Generation (AMG) refers to computational methods for composing music that reflects or evokes emotions, with applications in healthcare \cite{elliott2011relaxing, fujioka2012changes, stewart2019music}, co-creativity, and entertainment \cite{gorini2011role, hutchings2019adaptive}. Prior work categorizes AMG approaches into rule-based, data-driven, optimization-based, and hybrid systems~\cite{dash2024ai}. Rule-based methods rely on predefined mappings between musical features and emotional states \cite{wallis2011rule, ehrlich2019closed}, whereas data-driven approaches learn such mappings from data using deep learning models. Recent transformer-based architectures \cite{vaswani2017attention} have greatly improved the ability to model long-term temporal structure and musical coherence, outperforming earlier Markov chain and LSTM-based methods. 
Building on these advances, several transformer-based systems have been developed for symbolic music generation \cite{huang2018music, dhariwal2020jukebox, bhandari2025text2midi}, demonstrating the capacity of self-attention to capture long-range harmonic and rhythmic patterns.
Among symbolic systems, MINUET \cite{maniktala2020minuet} generates sentence-level, mood-conditioned music from text using a Markov chain trained on chord transitions, while SentiMozart \cite{madhok2018sentimozart} and Bardo Composer \cite{ferreira2020computer} use affective cues such as facial expressions or speech. More recently, audio-based text-to-music models such as MusicLM \cite{agostinelli2023musiclm} and Noise2Music \cite{huang2023noise2music} employ diffusion or auto-regressive transformers guided by large language models to produce music from short textual prompts. However, these systems are primarily conditioned on short descriptive text that characterizes the desired musical output (e.g., “calm piano melody” or “energetic electronic beat”) rather than on narrative sentences. Consequently, they do not attempt to model emotions emerging from longer, story-like text, which is the central goal of our work.

Several conceptual models have been proposed in the research literature to categorize and interpret emotions. The most popular models in NLP literature have been the categorical model and the dimensional model \cite{plaza2024emotion}. In the categorical model, emotions are discretized and assigned to separate classes, as in Ekman’s framework of six basic emotions—happiness, sadness, fear, anger, surprise, and disgust \cite{ekman1969repertoire}. The dimensional model, in contrast, represents emotions as continuous variables along a set of psychological axes. The most widely adopted formulation is the valence–arousal–dominance (VAD) model \cite{russell1977evidence}, where valence measures the positivity or negativity of an emotion, arousal measures its level of activation or intensity, and dominance captures the degree of control or power associated with it.

In many affective computing and music emotion recognition studies, emotions are projected onto the two-dimensional valence–arousal (VA) plane, which simplifies the VAD model by focusing on the most perceptually salient dimensions. This plane provides an intuitive framework where emotions such as joy and anger both have high arousal but differ in valence, while sadness and calmness share low arousal but differ in valence. The VA representation offers a continuous and interpretable space that facilitates mapping between textual and musical affect, and it serves as the foundation for the emotion conditioning in our work.

\section{Datasets}
Since no existing dataset directly links textual emotions with corresponding musical emotions, we construct one by merging two complementary resources: GoEmotions \cite{demszky2020goemotions}, a large-scale corpus of text annotated with fine-grained emotion labels, and EMOPIA \cite{hung2021emopia}, a symbolic music dataset categorized by emotional quadrants. This merger allows us to create aligned text–music pairs that share similar affective dimensions, enabling cross-modal learning of emotional expression between language and music.

\textbf{GoEmotions.} To represent the emotional content of narrative-like texts, we use the GoEmotions dataset, which contains over 58k English Reddit comments manually annotated for 27 emotion categories or neutral. The comments are sampled from 482 subreddits, each with at least 10k comments, ensuring broad topical diversity and a rich distribution of emotional expressions.

\textbf{EMOPIA.} Research literature on Affective Music Generation, lacks large-scale human labeled MIDI datasets. Most  available datasets are small and contain less than 200 MIDI files paired with emotion labels \cite{ferreira2021learning, panda2013multi}. We use the largest dataset available, EMOPIA \cite{hung2021emopia}, which contains 1,087 music clips from 387 songs and annotated for emotion by 4 human annotators. There are 4 unique emotion labels in this dataset, each corresponding to different quadrants in the valence-arousal plane. More information on EMOPIA is provided in the Appendix (Section~\ref{sec:quad_distribution}).

\textbf{GiantMIDI-Piano.} The third dataset used in this study is GiantMIDI-Piano \cite{kong2020giantmidi} which was constructed for various research areas such as classical music analysis and symbolic music generation. It contains 10,855 solo piano works composed by 2,786 different composers.

\subsection{Creating the Story2MIDI Dataset}
\begin{figure}
    \centering
    \includegraphics[width=1\linewidth]{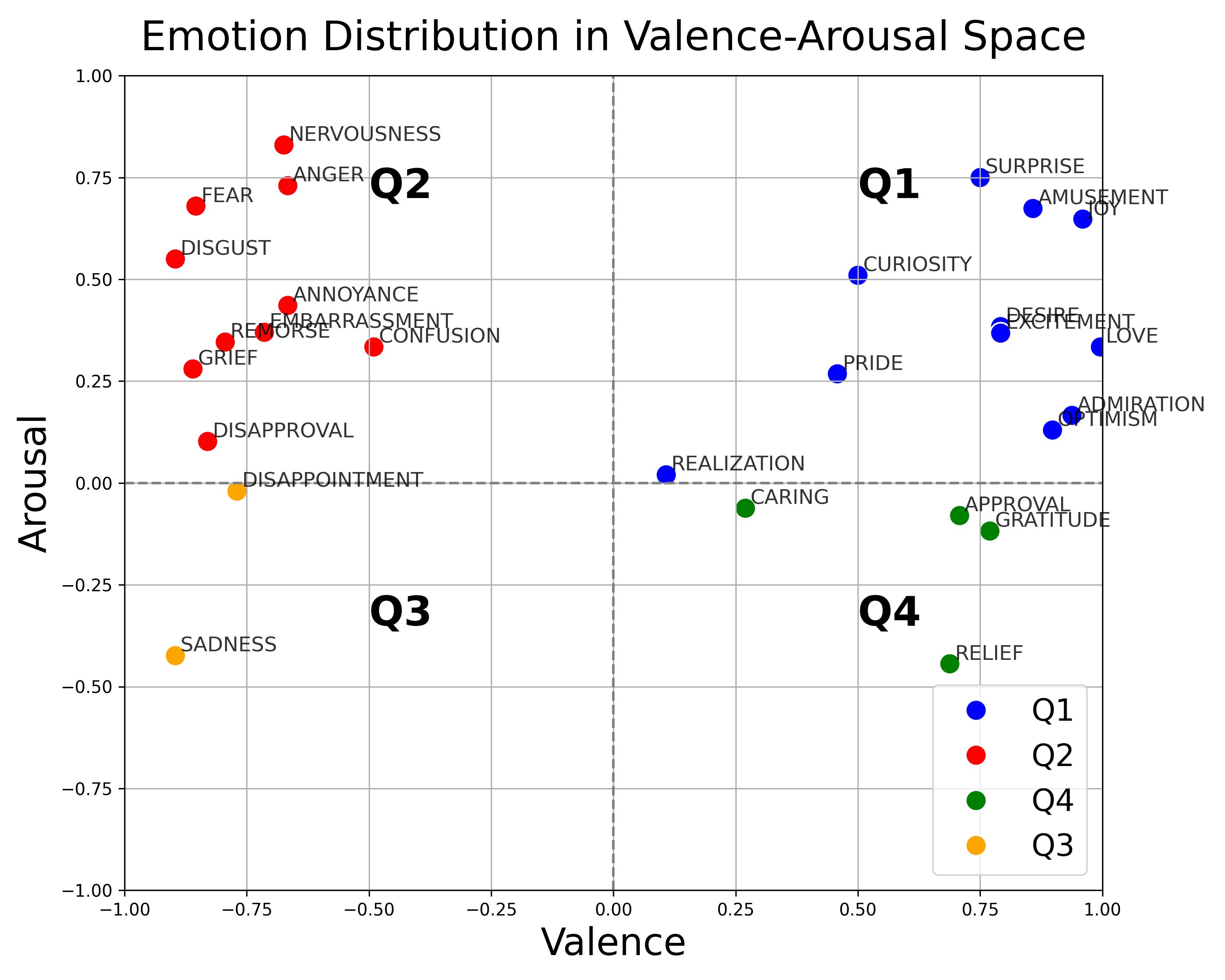}
    \caption{27 different emotion classes from GoEmotions dataset located on the AV plane. Arousal and valence scores were looked up from the NRC VAD lexicon.}
    \label{fig:emotions_va}
\end{figure}

EMOPIA includes four unique emotion labels $\{Q_1, Q_2, Q_3, Q_4\}$ for its music files, each corresponding to a quadrant of the arousal-valence (AV) plane (see Figure \ref{fig:emotions_va}).
The AV plane refers to Russell's Circumplex model of affect, which is a well-known framework for representing emotion \cite{russell1977evidence}.
It represents affect graphically across two dimensions: arousal (from deactivation to activation) on the vertical axis and valence (from unpleasant to pleasant) on the horizontal axis.
Specifically, $Q_1$, $Q_2$, $Q_3$, and $Q_4$ correspond to \textit{high arousal–high valence} (e.g., joy, excitement), \textit{high arousal–low valence} (e.g., anger, fear), \textit{low arousal–low valence} (e.g., sadness, disappointment), and \textit{low arousal–high valence} (e.g., relief, gratitude), respectively.

In order to pair our text data with EMOPIA files, we categorized the 27 emotion labels to four AV quadrants. To do so, we used the NRC VAD lexicon \cite{mohammad2018obtaining}, a large lexicon of human-annotated arousal and valance scores for 55k words. Using this lexicon, we assigned one of the four quadrant labels to each emotion (constructing GoEmotionsAV). The resulting mappings and their positions on the AV plane are shown in Figure \ref{fig:emotions_va}. (Appendix ~\ref{sec:vad_to_quad} provides more details on the VAD and the mapping to EMOPIA.)

\subsection{Music Representation}


We use the Musical Instrument Digital Interface (MIDI) as our format for encoding and generating symbolic music data because of its capacity to encode musical performance data through discrete parametric events. It is also a compact representation that is still expressive enough to convey emotion-related aspects of music. There is also the advantage of the numerous MIDI datasets that are publicly available to use as training data. In our training, we view each MIDI file in our datasets as a sequence of discrete musical events, enabling the use of sequence-to-sequence models for effective learning and generation.

To tokenize MIDI files before feeding them into our models, we employ the REMI tokenizer \cite{huang2020pop} from the MIDITok library\footnote{\href{https://miditok.readthedocs.io/en/latest/}{See MIDITok's documentation.}}.  REMI represents a music score as a sequence of \textit{pitch}, \textit{velocity}, and \textit{duration} tokens, supplemented with additional tokens such as \textit{bar} and \textit{position} to encode temporal structure. In REMI tokenization, each token family (e.g., pitch) has a limited set of possible values, resulting in a relatively small overall vocabulary. A Generative model trained with REMI must learn the correct sequential ordering of these token types to ensure that the generated output can be interpreted and played correctly by downstream MIDI players.

\section{Evaluation Approach}
Despite significant advances in generative modeling of sound and music, the evaluation of such systems lacks established methodology \cite{vinay2022evaluating}. Aside from listening studies, researchers have suggested a set of objective metrics for evaluating audio reconstructions and neural networks trained to classify the generated audio into existing categories. In this study, we employ both a listening study and a set of objective metrics used in previous studies for evaluating symbolic music generation. We generate 100 MIDI samples per quadrant for texts from our test set (GoEmotions) and report the metrics. 

\subsection{Preliminary Listening Study}
\label{sec:qualEval}
As part of our preliminary listening study, and to get an initial sense of our model's performance, we asked human listeners how well the generated music aligns with the emotion intended by the text. This setting is even more subjective than typical music generation tasks, as it involves two layers of interpretation: listeners must first perceive the emotional content of the story and then assess whether the accompanying music appropriately reflects that emotion. In the first phase, each participant was presented with a piece of text and asked to assign it to one of the four quadrants in the valence-arousal space. In the second, they were shown two music options corresponding to that story and asked to choose the one they believe best evoked the same emotion as the story. This two-step design allowed us to assess both the consistency of text-level emotion perception and the perceived emotional alignment between story and music.

Since not all candidate music options correspond to the exact quadrant label of the story selected by the participant, we evaluated emotional alignment based on the valence–arousal dimensions rather than exact quadrant matches. Each quadrant was mapped to a binary valence (positive/negative) and arousal (high/low) value, and for each participant response, we assessed whether the selected music sample matched the participant’s perceived emotion of the story along the valence and arousal axes. This yields valence accuracy, arousal accuracy, and full valence+arousal accuracy, offering a more flexible and perceptually grounded measure of emotional agreement in the forced-choice setting. 

\subsection{Objective Evaluation}
\label{sec:metrics}


To quantitatively assess the emotional characteristics of the generated music, we evaluate our model using objective metrics that reflect the two primary affective dimensions: valence and arousal. 
Following the methodology proposed in the original EMOPIA paper \cite{hung2021emopia}, we compute several music-theoretic and acoustic descriptors known to correlate with emotional expression, allowing us to compare our generated samples against the reference EMOPIA dataset along these dimensions.


\textbf{Valence-related metric.}
Prior work \cite{panda2020audio, livingstone2010changing} reports that music in a major key tends to evoke positive affect including happiness and serenity (Q1, Q4), while minor keys tend to be associated with negative emotions like sadness and anger (Q4 and Q2).
As in \cite{hung2021emopia}, we extract the overall key from each of our generated files using the Krumhansl-Kessler algorithm \cite{Krumhansl_2001} for key detection implemented in the music21 Python library\footnote{\url{https://www.music21.org/music21docs/}}. 
Then, we calculate the major key ratio of our generated files in each quadrant. That is, we calculate what proportion of files in each quadrant have a major key out of all the generated files for that quadrant.
We expect that files generated from stories in Q1 and Q4 should have a higher major key ratio, while files generated from stories in Q2 and Q3 should have lower major key ratio.

\textbf{Arousal-related metrics.} Again following from \cite{hung2021emopia}, we calculate note length and note velocity to evaluate the model's ability to evoke arousal-related musical features. Note length is the average duration of notes in seconds and note velocity is the speed with which a piano key is hit, and captures dynamics or loudness.
Note velocity and density are expected to be higher in music with higher arousal (Q1 and Q2) and lower in music with lower arousal (Q3 and Q4), while note length is higher in music with lower arousal~\cite{panda2020audio, livingstone2010changing}. 

Ideally, the distribution of these metrics across the quadrants should mirror the distributions seen in EMOPIA. We both visually compare the metrics of our generated files to those on the EMOPIA files and use t-tests to assess whether the generated data maintains the statistically significant difference in the metrics for the EMOPIA data between high and low valence and arousal.

\section{Modeling Approach}
\begin{figure*}[htpb]
    \centering
    \includegraphics[width=\textwidth]{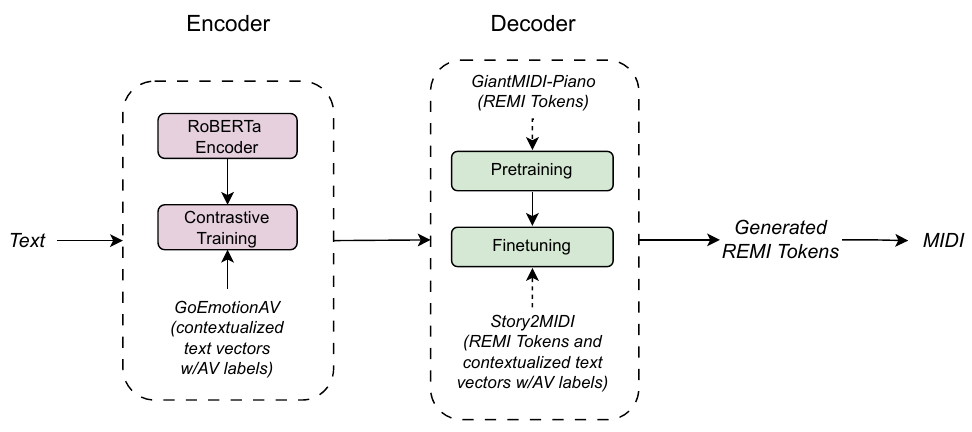}
    \caption{Overview of Story2MIDI. The encoder of our model is a RoBERTa encoder which is contrastively fine-tuned to distinguish emotions better. The decoder of our model is a transformer decoder with 3 decoder layers and four self-attention heads per decoder layer.} 
    \label{fig:system_figure}
\end{figure*}


To generate music conditioned on a specific emotion inferred from text, we employ an encoder–decoder Transformer architecture. The textual emotion signal is captured using a \texttt{RoBERTa-large} encoder \cite{liu2019roberta}, which is contrastively fine-tuned to better capture nuanced emotion differences across quadrants. The decoder is a Transformer-based module consisting of three layers, each with four attention heads, designed to model the temporal and harmonic structure of symbolic music. Figure \ref{fig:system_figure} shows the overview of our model architecture.

\subsection{Contrastive Training for the Encoder}
The encoder is initialized from \texttt{RoBERTa-large} and fine-tuned on the GoEmotions dataset, where the 27 emotion labels are replaced with their corresponding quadrant labels. Each batch contains examples from multiple quadrants; pairs within the same quadrant are treated as positive pairs, and pairs from different quadrants as negatives.

Let $\mathbf{h}_i$ denote the normalized embedding of example $i$, and $\mathcal{P}(i)$ denote the set of positives for $i$ in the batch. 
The supervised contrastive loss \cite{khosla2020supervised} is defined as:
\begin{equation}
\mathcal{L} = - \frac{1}{N} \sum_{i=1}^{N} \frac{1}{|\mathcal{P}(i)|} 
\sum_{p \in \mathcal{P}(i)} 
\log \frac{\exp(\mathbf{h}_i \cdot \mathbf{h}_p / \tau)}
{\sum_{a=1}^{N} \exp(\mathbf{h}_i \cdot \mathbf{h}_a / \tau)},
\label{eq:supcon}
\end{equation}
where $\tau$ is a temperature hyperparameter controlling the sharpness of similarity weighting.

This in-batch formulation enables efficient training with many negatives and produces an embedding space in which quadrants form separable clusters. 
The contrastively fine-tuned encoder is subsequently frozen and used as the text-conditioning module in our generative model.

For this contrastive training step, we unfreeze the final four transformer layers of the RoBERTa encoder and fine-tune for 500 epochs using the AdamW optimizer with a learning rate of $1e-5$ and a batch size of 16. We use a balanced subset of the GoEmotions dataset, excluding the portion reserved for pairing with EMOPIA. Because quadrant $Q_3$ contains the fewest samples, we downsample all quadrants to 3,582 examples each, resulting in a total of 14,328 text instances evenly distributed across the four emotion quadrants.

The pretrained RoBERTa encoder produces highly overlapping representations across the four emotion quadrants, whereas the contrastively fine-tuned encoder yields noticeably more separated clusters. This confirms that contrastive training strengthens the emotional structure of the embedding space, while still reflecting the inherently gradual and subjective nature of affective categories.
(See Figure~\ref{fig:contrastive_training} in the Appendix (Section~\ref{sec:contrastive_training} for the t-SNE visualizations of the story embeddings before and after contrastive training.) 

\subsection{Pre-training the Decoder}
EMOPIA contains only 1,078 MIDI samples, which is insufficient for a Transformer decoder to effectively learn the underlying dependencies between MIDI tokens. To address this, we pre-train the decoder on the larger GiantMIDI-Piano dataset. During pre-training, we input zero tensors in place of text embeddings and train the decoder using a causal language modeling objective, enabling it to learn the probability distribution over MIDI tokens in an auto-regressive manner.

We use a learning rate of $1\mathrm{e}{-4}$, a batch size of 64, and train the model using the Adam optimizer \cite{kinga2015method}, with 1,000 warm-up steps in the learning rate scheduler. The decoder is trained for 300 epochs on an NVIDIA L40 GPU, totaling approximately 75 GPU hours. The loss steadily decreases for both training and validation sets, with the rate of improvement slowing noticeably after approximately 200 epochs. 

To evaluate the quality of pre-training, we examine how closely the token type distributions in the generated samples align with those in the original pre-train data. Specifically, we compute the average frequency of each token type across tracks in the pre-train dataset and compare these with the frequencies observed in generated outputs at various training checkpoints. The token distributions from checkpoints at 250 epochs most closely resemble the original distribution. Notably, despite the training and validation losses continuing to decrease through 300 epochs, the generated samples exhibit greater divergence from the pre-train token distribution. Based on this observation, we select the checkpoint at 250 epochs for fine-tuning on EMOPIA, as it achieves the best balance between loss minimization and distributional fidelity to the pre-training data.

\subsection{Fine-tuning on EMOPIA}
After pre-training the decoder, it is capable of generating coherent symbolic music sequences. 
To align the generation process with emotion, we fine-tune the full model on our constructed dataset, which pairs short text blurbs with affectively labeled musical pieces from EMOPIA. 
The text encoder provides a strong affective signal that conditions the decoder during this stage. To prevent catastrophic forgetting \cite{kirkpatrick2017overcoming}, all decoder layers except the final one are frozen.
This strategy allows the model to adapt to emotion–music associations while preserving the musical structure learned during pre-training.
We also experimented with unfreezing the token embedding layer but observed irregular note repetitions in the generated outputs.
Therefore, we retain all other parameters frozen and fine-tune only the final decoder layer with a learning rate of $1e-5$ for 300 epochs. We saved model checkpoints every 20 epochs and selected the best-performing epoch based on the objective valence and arousal metrics of the generated samples, as described in Section~\ref{sec:metrics}.
We define “best performance” as the checkpoint that minimizes the Mean Squared Error between the average valence–arousal metrics of each quadrant in the generated data and those of the corresponding EMOPIA samples.

\subsection{Decoder Validation}

To validate that the decoder could generate MIDI based on AV quadrants, we conducted an ablation experiment where, instead of using text embeddings generated by the encoder, we directly encoded each quadrant label (e.g., ``quadrant\_1'') as the input to the decoder. This simplified setup produced MIDI samples whose valence–arousal metrics closely matched EMOPIA on all of the metrics described in Section~\ref{sec:metrics}.

\section{Experiment Results}
\begin{figure*}[htpb]
    \centering
    \includegraphics[width=0.86\textwidth]{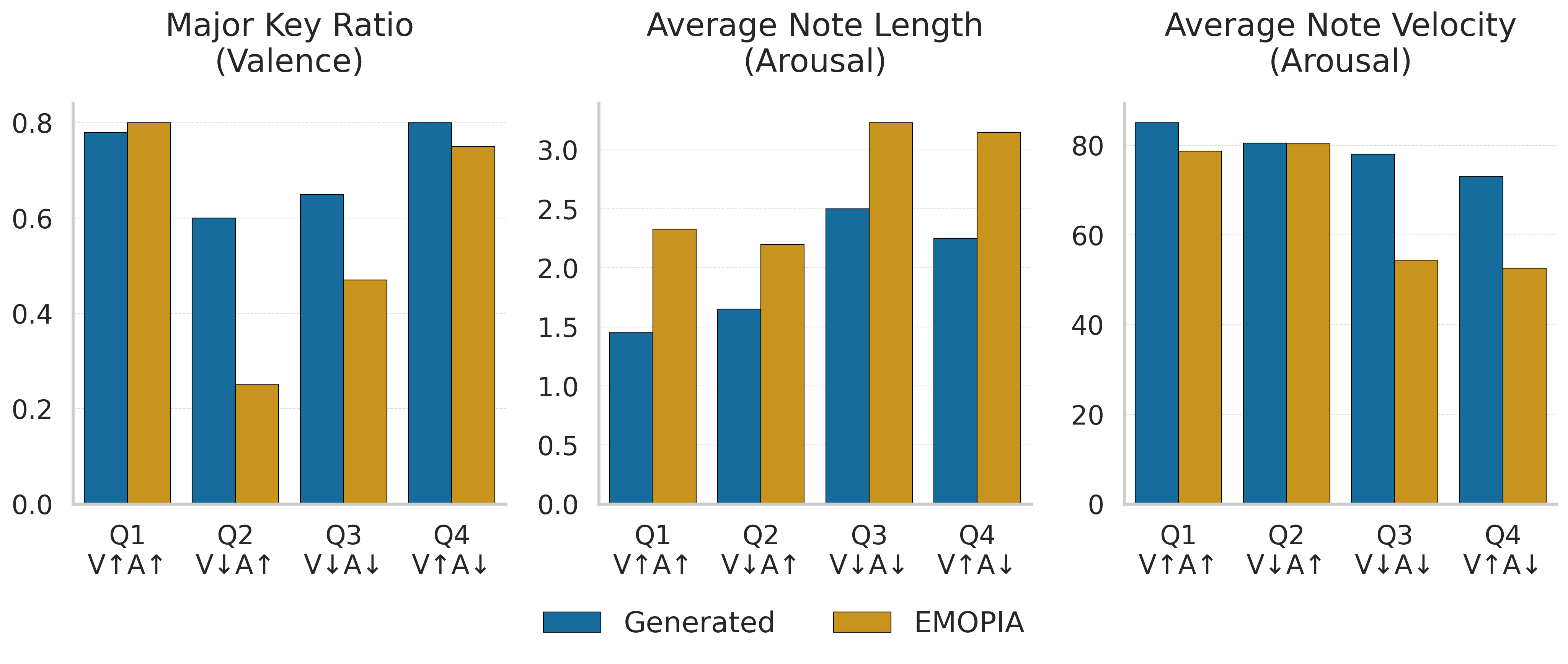}
    \caption{
    Valence-arousal metrics comparing our generated samples after 300 epochs of fine-tuning with the EMOPIA dataset. 
    The plots show differences in average note length, note velocity, and major key ratio across the four emotion quadrants (Q1–Q4). 
    In the Major Key Ratio plot, generated samples exhibit a statistically significant difference between positive-valence (Q1–Q2) and negative-valence (Q3–Q4) quadrants, 
    reflecting valence sensitivity in our model. 
    Similarly, the Average Note Length plot shows a significant difference between high-arousal and low-arousal quadrants, 
    indicating the model’s ability to capture arousal variation through musical structure. 
    }
    \label{fig:metrics_figure}
\end{figure*}

\subsection{Listening Study}
\label{Listening Study}
Using the methodology described in Section \ref{sec:qualEval}, we conducted a preliminary listening study with four participants. Table~\ref{tab:subj_eval} shows the results of our listening study. Our listeners rate the model output as better fitting the arousal (0.70 accuracy) dimension than either valence (0.53 accuracy) or valence+arousal (0.40 accuracy). The chance levels for valence, arousal, and combined valence–arousal predictions are 0.5, 0.5, and 0.25, respectively. As shown in Table \ref{tab:subj_eval}, the model’s generated music exceeds these baselines across all dimensions, particularly for arousal (0.70) and joint valence–arousal alignment (0.40). This indicates that participants perceived a correspondence between the emotional tone of the text and the music beyond random expectation. A larger participant pool would provide more reliable estimates of perceptual alignment and allow for statistical significance testing. Expanding the study to include a more diverse set of listeners is an important direction for future work and will help validate the generalizability of these preliminary findings.

\begin{table}[h]
\caption{Subjective evaluation of perceived emotional alignment between story and music using valence and arousal dimensions.}
\centering
    \begin{tabular}{lccc}
    \toprule
    \textbf{Dimension} & \textbf{Accuracy} \\
    \midrule
    Valence     & 0.53 \\
    Arousal     & 0.70 \\
    Valence+Arousal    & 0.40 \\
    \bottomrule
    \end{tabular}
\label{tab:subj_eval}
\end{table}

\subsection{Objective Evaluation}

Figure \ref{fig:metrics_figure} depicts the arousal (note length and note velocity) and valence (major key ratio) metrics from our generated samples compared against EMOPIA MIDI files. Major key ratio reflects valence, while average note length and note velocity reflect arousal.

\textbf{Valence-related metric.} 

For each generated MIDI file, we compute the major key ratio and conduct an independent-samples 
t-test comparing positive-valence (Q1 \& Q4) versus negative-valence (Q2 \& Q3) quadrants.
The difference is statistically significant (\emph{p} = 0.026), indicating that the model aligns harmonic mode with emotional valence.



\textbf{Arousal-related metrics.} The results of a t-test show that our generated samples exhibit a statistically significant difference in average note length between high-arousal (Q1 \& Q2) and low-arousal (Q3 \& Q4) quadrants (\emph{p} $<$ 0.01), indicating that the model successfully captures temporal variation associated with musical energy. In contrast, differences in average note velocity are not significant across quadrants. Nevertheless, higher-arousal quadrants (Q1 \& Q2) still exhibit a trend of higher average note velocities in our generated samples, consistent with the expected relationship between arousal and musical intensity.

\subsection{Discussion}

For the valence metric, across three quadrants the major-key ratio of our generated samples tracks EMOPIA reasonably well. The largest divergence is Q2, where our model produces a much higher major-key rate than EMOPIA. This likely reflects a major-key bias caused by the dataset having more major-key tracks than minor-key ones, combined with a limited fine-tuning set.

For the arousal metrics, the fine-tuned model clearly captures the note-length pattern: shorter notes in high-arousal quadrants (Q1 \& Q2) and longer notes in low-arousal quadrants (Q3 \& Q4), closely matching EMOPIA. The differences in average note velocity are less pronounced across quadrants, likely because REMI quantizes MIDI velocities (0–127) into a limited set of discrete bins, which compresses the dynamic range and reduces between-quadrant separability.

The quantitative results partially mirror the findings of the listening test in so far as the model aligning at least some of the musical characteristics with arousal-valence text-based characteristics.  The qualitative results indicate that the model is doing a better job of capturing arousal than valence, with the valence results being close to chance. This could be due to our model's tendency to overgenerate in the major. It also suggests that a larger number of quantitative metrics would be useful to assess where the valence aspects of generation may be lacking.

\section{Conclusions}
In this work, we introduced Story2MIDI, a transformer-based model for generating emotionally aligned symbolic music from textual input. To enable this task, we constructed a paired text–music dataset and developed an encoder–decoder framework designed to capture affective correspondence between language and music. The encoder was contrastively fine-tuned to produce distinctive emotion representations across valence–arousal quadrants, while the decoder was first pretrained on a large corpus of symbolic music and subsequently fine-tuned on a smaller, emotion-labeled dataset for emotion-conditioned music generation. We evaluated Story2MIDI using both subjective listener judgments and objective musical metrics; together, these evaluations demonstrate the model’s effectiveness and highlight several challenges that merit further exploration


Overall, our model effectively captures affective musical features associated with emotional expression. In particular, we observed a statistically significant difference in average note length across emotion quadrants, suggesting that the model adjusts rhythmic structure to reflect variations in emotional intensity. In contrast, note velocity did not show a significant difference, indicating that dynamic intensity was not as strongly influenced by the conditioning signal. These results suggest that the model relies more on temporal aspects of music to convey affect, and that incorporating additional expressive features or larger training data could further enhance emotional alignment. 

Another avenue of improvement is in the text-to-emotion mapping. Our ablation study examining the performance of the model when inputting emotional quadrants labels, rather than text, into the decoder demonstrates that the emotion-to-music mapping component of the architecture is robust. Thus, while our model can extract a meaningful affective representation from natural language text, a purely symbolic quadrant embedding still provides a stronger and less noisy conditioning signal. And finally, as discussed in Section \ref{Listening Study}, we plan to increase the scale of the listener study to validate our initial findings.

\section*{Acknowledgment}
This work is supported by the National Science Foundation (NSF) award 2228910.

\bibliographystyle{IEEEbib}
\bibliography{refs}

\appendix
\subsection{Obtaining Quadrants for GoEmotions Categories}
\label{sec:vad_to_quad}
To assign each of the 27 GoEmotions categories to an emotion quadrant, we rely on a Valence–Arousal–Dominance (VAD) lexicon. A VAD lexicon provides normative ratings for words along three continuous affective dimensions: valence (pleasantness vs.~unpleasantness), arousal (activation or intensity), and dominance (sense of control). These scores, typically ranging from 0 to 1, are derived from large-scale human annotation studies and are widely used in affective computing as a quantitative representation of emotional meaning.
For our purposes, we use the valence and arousal dimensions to determine each emotion category’s location in the two-dimensional affect space. For each of the 27 categories, we lookup the corresponding entry in the VAD lexicon and extract its valence and arousal scores. We then assign the category to one of Russell’s four quadrants (high/low valence × high/low arousal). For example, if a category has valence = 0.54 and arousal = 0.43, it falls into the high-valence, high-arousal region and is therefore assigned to Quadrant 1. This mapping provides a consistent, interpretable way to align textual emotion labels with the valence–arousal structure used for conditioning the music generation model.

\subsection{Quadrant Balance Analysis}
\label{sec:quad_distribution}
This section reports the full distribution of text–MIDI pairs across the four valence–arousal quadrants. Table~\ref{tab:quad-dist} summarizes the number of samples in each quadrant used in our experiments.

\begin{table}[h]
    \centering
    \caption{Distribution of samples across quadrants in the Story2Music dataset (based on EMOPIA).}
    \begin{tabular}{lcc}
    \toprule
    \textbf{Quadrant} & \textbf{Description} & \textbf{Samples} \\
    \midrule
    Q1 & High Valence, High Arousal & 250 \\
    Q2 & Low Valence, High Arousal  & 265 \\
    Q3 & Low Valence, Low Arousal   & 253 \\
    Q4 & High Valence, Low Arousal  & 310 \\
    \bottomrule
    \end{tabular}
    
    \label{tab:quad-dist}
\end{table}

As shown, all quadrants are represented with comparable counts. Because our model is fine-tuned in a conditional generation setup rather than a multi-class classification setup, minor differences of this scale do not impose strong bias toward any particular quadrant.

\subsection{Contrastive Training}
\label{sec:contrastive_training}
\begin{figure}
    \centering
    \includegraphics[width=0.8\linewidth]{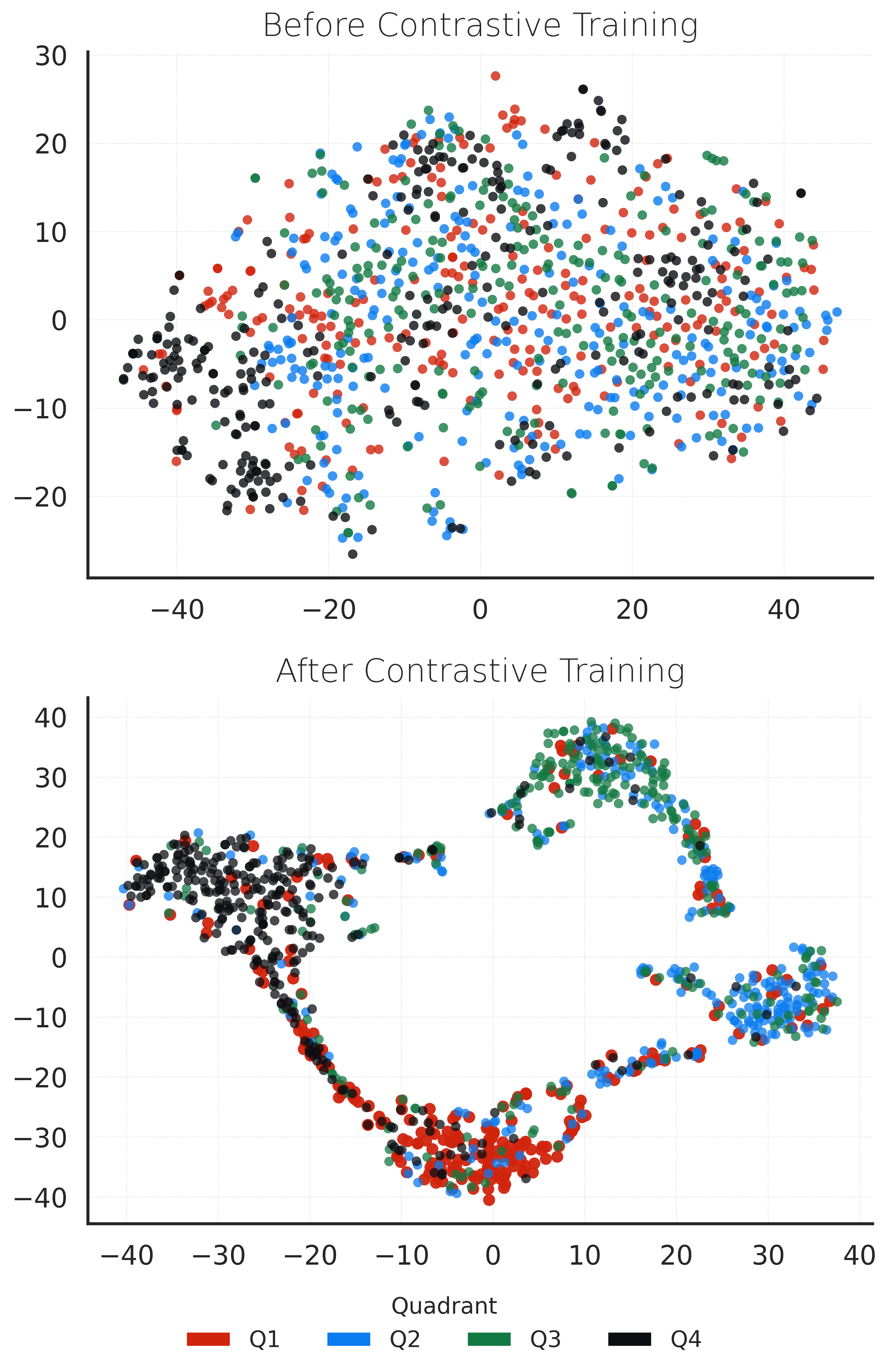}
    \caption{t-SNE visualization of story embeddings before and after contrastive training. Contrastive fine-tuning leads to more distinct quadrant clusters.}
    \label{fig:contrastive_training}
\end{figure}

As seen in Figure~\ref{fig:contrastive_training}, the initial representations are highly overlapping across emotion quadrants, indicating that the pretrained  RoBERTa model does not distinctly separate texts by their emotional valence–arousal characteristics. This reflects the nuanced and subjective nature of emotion in text, where semantically diverse pieces of text may evoke similar affective responses, and emotion boundaries are inherently fuzzy.

The bottom plot shows embeddings from our contrastively fine-tuned encoder. Here, the clusters corresponding to the four quadrants become more distinct, which demonstrates that contrastive training encourages the encoder to map affectively similar stories closer together in the embedding space, thereby strengthening the emotional signal for downstream generation. Nevertheless, some interleaving among the clusters persists, underscoring the complexity and subjectivity of emotion perception.

\end{document}